\documentclass[conference,10pt]{IEEEtran}
\IEEEoverridecommandlockouts
\usepackage{cite}
\usepackage{amsmath,amssymb,amsfonts}
\usepackage{algorithmic}
\usepackage{graphicx}
\usepackage{textcomp}
\usepackage{xcolor}

\usepackage[linesnumbered,ruled]{algorithm2e}
\usepackage{array}
\usepackage[caption=false,font=scriptsize,labelfont=sf,textfont=sf]{subfig}
\usepackage{stfloats}
\usepackage{url}
\usepackage{color}
\usepackage{verbatim}
\usepackage{tasks}
\usepackage{authblk}

\def\BibTeX{{\rm B\kern-.05em{\sc i\kern-.025em b}\kern-.08em
    T\kern-.1667em\lower.7ex\hbox{E}\kern-.125emX}}

\begin{document}


\title{A Queueing Theoretic Perspective on Low-Latency LLM Inference with Variable Token Length}



\author[1]{Yuqing Yang}
\author[2]{Lei Jiao}
\author[1]{Yuedong Xu$^{\ast}$\thanks{*Corresponding author.}\thanks{This work was supported in part by the National Natural Science Foundation of China under Grant 62072117, the Natural Science Foundation of Shanghai under Grant 22ZR1407000, and the U.S. National Science Foundation under Grants CNS-2047719 and CNS-2225949.}}
\affil[1]{School of Information Science and Technology, Fudan University, China}
\affil[2]{Center for Cyber Security and Privacy, University of Oregon, USA}

\maketitle

\begin{abstract}

Large language models (LLMs) propel the prosperity of interactive AI applications showcased by ChatGPT that demand timely response of inference services. However, LLM inference is computation intensive and memory intensive, and improper parameter configuration at LLM platforms may exacerbate the inference time. In this paper, we analyze the impact of LLM output token distribution on the inference queueing delay, where the max-token clipping and the batched inference are considered. By formulating an M/G/1 model, we observe that enforcing a maximum output token limit on a very small fraction of inference requests can significantly reduce the queueing delay, and our model facilitates the selection of the optimal limit. For the batch inference, we model the service process as a bulk queue in which the batch processing time is affected by the batch size and the maximum token size inside this batch jointly. The queueing delays of the batching of all buffered requests (dynamic batching), the batching of constant number of requests (fixed batching), and the batching without intra-batch waiting (elastic batching) are derived. Experimental results show that our mathematical models coincide with the event-driven simulations well.

\end{abstract}


\section{Introduction}
\label{sec:introduction}
A Large Language Model (LLM) is a gigantic neural network trained on massive amount of text data such as GPT \cite{37} and LLaMA \cite{3}. It is not only capable of generating natural language sentences, but also possesses the power of understanding textual meaning. Nowadays, LLMs have been comprehensively applied in almost every aspect of content generation, and rapidly expand to search engine and software engineering. It is even believed that LLMs even light the way toward artificial general intelligence.
As a generative model, the way that a LLM creates content is called ``inference''. The LLM resides at one or more computing nodes, and users submit their inference requests to the LLM platform for processing. 
Intuitively, an input or output request with larger token length demands more time,
thus affecting the latency of LLM inference.

Recently, there have been a lot of efforts to improve LLM inference latency concerning token length. One approach to reducing decode latency is to enforce a maximum output token limit. However, a large token limit can still result in significant queuing delays when inference requests arrive at the LLM platform and are processed on a first-come-first-serve (FCFS) basis. Conversely, a shorter token limit may impair inference quality. To alleviate prefill overhead, batch inference is utilized to compute the KV matrices of multiple requests simultaneously. Due to the randomness in the batched requests, their input and output token lengths are usually misaligned. Consequently, the token lengths of all requests are padded to match the maximum length of the current batch before being fed into the self-attention module, causing that their inference times are uniform. The choice of batch size affects both the waiting time for requests to be grouped and the batch inference time. A more sophisticated batching technique include continuous batching \cite{36}. In summary, the randomness in token length is a crucial factor in providing low-latency LLM inference services. However, there is currently no queuing theoretic analysis on the end-to-end service delay.

In this paper, we first explore how the distribution of token lengths, particularly the output token length, impacts queueing delay at both the decode and prefill stage, and what are potential measures for improving overall LLM service quality. Observing that the inference latency of an individual request is proportional to its output token length, we model the service process as an M/G/1 queue \cite{21} and derive the queuing delay in closed form. An intuitive finding is that the heavy tail of output token length in a few requests significantly extends the average queuing delay. This can cause a considerable percentage of impatient users to leave the LLM platform before their requests are processed. We propose configuring an appropriate maximum output token limit, as a slight reduction in inference quality for a very small percentage of requests can significantly decrease the average queuing time. 

We next investigate the mathematical model of queuing delay for batch inference. Our focus is placed on the more complex dynamic batching \cite{18} where a GPU processes all buffered requests in one go. In this context, the batch size is uncertain, and the inference latency of a batch depends not only on the batch size but also on the largest output token length among all the requests in the batch. We model this dynamic batching service process with an unbounded batch size as an M/G/1 queue, where the service time distribution is correlated with both the arrival rate and the output token length distribution. We explicitly derive a tight upper bound for the average queuing delay. Our observations indicate that when the output token length follows a heavy-tailed distribution, setting an optimal maximum batch size is beneficial, especially in trading off waiting time for faster batch inference. To find this optimal value, we derive an alternative model for static batching \cite{24}, where the batch size is constant, and compute the optimal value as a function of the arrival rate and the output token length distribution. Applying this optimal maximum batch size to dynamic batching, we observe a considerable decrease in the queuing delay for dynamic batching.
In the realm of traditional dynamic batching techniques, there exists an inherent issue where shorter responses are compelled to wait for the completion of longer ones. This scenario inevitably leads to an increase in queueing delays. To address this, we model the system that enables replies generating fewer tokens to be expedited back to users without the need for padding. This elastic batching ensures a minimized queueing delay, irrespective of the distribution that the output token size adheres to.

\section{Background and Motivation}
\label{sec:background}
In this section, we briefly introduce the inference procedure of Transformer-based large language models, and show two factors that affect the LLMs inference latency.

\subsection{LLM Inference Basics}

\begin{figure}[!t]
\centering
\subfloat[LLaMA-2 architecture]{\includegraphics[width=1.9in]{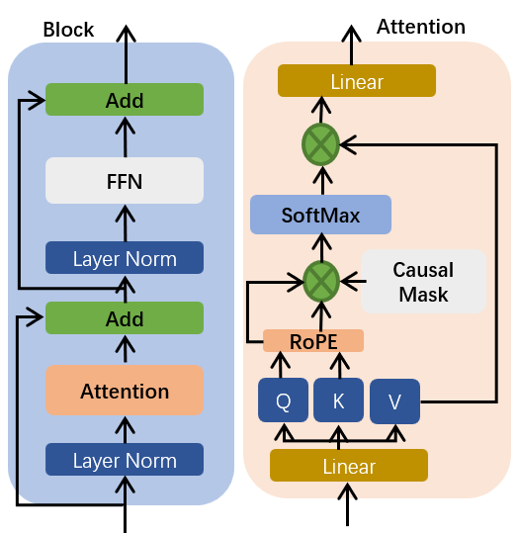}%
\label{fig2.1.1}}
\hfil
\subfloat[KV cache]{\includegraphics[width=1.5in]{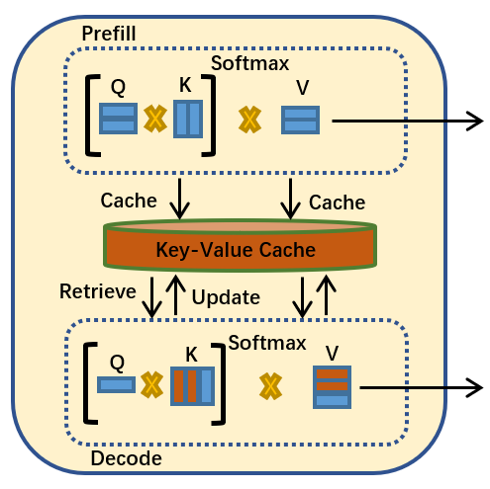}%
\label{fig2.1.2}}
\caption{LLM inference basics}
\label{fig2.1}
\end{figure}

We take LLaMA-2 as an example model for general LLMs. The key component of LLaMA consists of a stack of blocks similar to the Transformer decoder blocks, as shown in Fig. \ref{fig2.1.1}. In a LLaMA block, the attention module distinguishes itself from a convolutional module. 
Each input token is derived with three values: query, key, and value. The model computes the dot product between the new query and the keys of preceding tokens to assess the relevance of prior tokens from the perspective of the current token. 
Subsequently, it utilizes Softmax on the dot products to generate weights, and computes the output as a weighted sum of the values based on these weights.

To generate a new token, 
the inference process is divided into two stages, as shown in Fig. \ref{fig2.1.2} \cite{19}.
The input consists of two tokens. Freshly computed tensors are depicted by blue grids, while reused tensors from the key-value cache are represented by red grids. 
In the prefill phase, which processes all the input tokens within a batch simultaneously, 
the keys and values need to be calculated and saved for each block and the cache will be filled when the first token is output. 
The decode phase that generates the second token to the last one performs the same operations as prefill, but only for the single token which was generated in the last autoregressive iteration.
In this phase, LLaMA only needs to compute the query, key, and value of the newly generated token. The key-value cache is employed and adjusted iteratively to generate tokens incrementally, so that the inference time to generate the first token is larger.

The optimization of LLMs inference latency can be broadly categorized into two types: one involves optimizing inference on a per-request basis, handling each request sequentially, while the other involves system-level optimizations for concurrent inference of multiple requests, as detailed in the following discussion.

\subsection{Impact of Tokens on Inference latency}

Both input and output tokens must undergo the inference and computation processes of large language models (LLMs), impacting the model's inference latency.
We represent the workload of an inference task as the number of tokens to be generated, and measure the inference latency with regard to different numbers of tokens. Consider that a \emph{LLaMA-2-7b-chat} model runs on an NVIDIA A100 GPU. Table \ref{tab:table2.1} shows the inference latency when the pair of input and output tokens change.
In each experiment, we collect 100 conversation requests with the LLM, record the generated inference latency, and compute their average values. 

\begin{figure}[!t]
\centering
\subfloat[]{\includegraphics[width=1.7in]{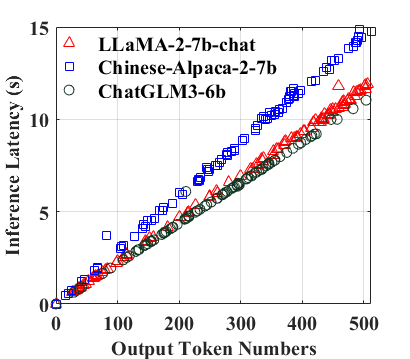}%
\label{fig2.3.1}}
\hfil
\subfloat[]{\includegraphics[width=1.7in]{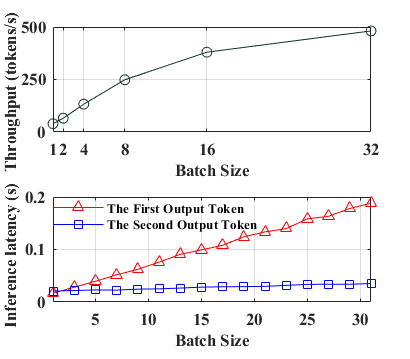}%
\label{fig2.3.2}}
\caption{Inference latency and throughput measurement on NVIDIA A100}
\label{fig2.3}
\end{figure}

One can observe that the inference latency is gently affected by the input token size, so we focus on the effect of output token size on the inference latency. We conduct the curve fitting in Fig. \ref{fig2.3.1} on Instruction-in-Wild \cite{35} dataset. We randomly sampled 100 instances for each model from those generating fewer than 512 output tokens and observed a linear relationship between inference latency and output token size. The observed experimental phenomenon can be explained by the fact that the number of input tokens primarily affects the time required to generate the first output token. When the number of output tokens is sufficiently large, the impact of input tokens on the overall inference latency becomes negligible. When generating the second output token till the last one, LLMs only need to compute the query, key and value of the newly generated token in an autoregressive manner and then perform some similar operations, so that their inference latency are approximatively identical. 

In practical LLMs services, a maximum token limit is enforced to avoid the demands of overlong output tokens. However, there lacks of a rigorous analysis on how this token limit influences the service latency. If the token limit is too small, it can hinder the generation of high-quality text. Conversely, if the token limit is too large, it can significantly increase the user's waiting time, potentially causing impatient users to abandon the system.

\begin{table}[!t]
\caption{the inference time with regard to different numbers of tokens
\label{tab:table2.1}}
\centering
\begin{tabular}{|c c | c c|}
\hline
Input/Output Tokens & Time (s) & Input/Output Tokens & Time (s)\\
\hline
(128, 128) & 2.91 & (64, 512) & 12.18\\
(128, 256) & 5.88 & (128, 512) & 12.63\\
(128, 512) & 12.63 & (256, 512) & 12.96\\
(128, 1024) & 23.47 & (512, 512) & 13.19\\
\hline
\end{tabular}
\end{table}

\subsection{Impact of Batching Inference}

Batch processing is an important feature of GPU inference. When processing a single inference request each time, the bottleneck that throttles the inference speed is the data loading instead of its computation. Therefore, multiple inference requests can be processed simultaneously so that the total I/O latency is amortized, and the GPU computational capacity is better utilized. Meanwhile, the decode phase is memory-bound due to the KV-cache footprint per request. For instance, we can only fit a maximum batch size of 49 requests at a sequence length of 1280 (the input token size is 256 and the output token size is 1024) for the LLaMA-2-7b-chat on an A100 GPU.
Within the feasible batch size range, the top half of Fig. \ref{fig2.3.2} illustrates the throughput of the inference stage for various batch sizes ($B$), with an input token size of 256 and an output token size of 1024. We observe that the throughput increases as the batch size grows.

We illustrate the token generation times of LLMs with regard to the batch size when fix the input token size (64) in the bottom half of Fig. \ref{fig2.3.2}. The generation times of all the tokens increase linearly with the batch size, where those of the first token are more sensitive to the batch size. The reason is that when generating the first token, all the key-value tensors of the input tokens are computed and cached, which increases the inference latency. LLMs input a vector whose size is input dimension times the batch size, and compute the output through a predefined sequence of operations, so that the generation times of the first token increases linearly with the batch size. When generating the second output token, LLMs only need to compute the query, key, and value of the newly generated token and then perform some similar operations, which also leads to a vector whose size depends on the batch size. For the production of additional tokens, the inference latency is similar to the second one within a feasible range of batch sizes. 

However, the number of output tokens obtained per request varies greatly, resulting in different distributions. As a result, different batching strategies are required to minimize inference latency effectively.

\section{Queuing analysis of Max-limit Clipping}

In this section, we formulate a probabilistic model to understand the impact of the maximum token limit on the inference service delay, and explore the optimal token limit. 

\subsection{$M/G/1$ Model}


Our goal is to obtain an explicit expression of the LLM inference delay. Without loss of generality, we model the LLM service process as a $M/G/1$ queue, where $M$ indicates that the requests arriving at the server follow the Poisson distribution, $G$ indicates that the service time of the server is generally distributed and $1$ indicates a single server. The inference requests are processed according to the first-come-first-served (FCFS) principle. Suppose that $\lambda$ is the mean arrival rate of requests and $\mu$ is the mean service rate of the server. Let $S$ denote the random variable for the serving time of an inference request, and let $\rho:=\frac{\lambda}{\mu}$. Denote by $W$ the queueing delay. When $\rho<1$, the queue is deemed as stable. 

According to \cite{21}, the mean queueing delay of a request is given by:
\begin{equation}
\label{eq:2.1}
E[W]=\frac{\lambda E[S^{2}]}{2(1-\rho)}.
\end{equation}

Each user has distinct requirements for the number of output tokens in their response. We denote by $n_{req}$ the number of output tokens that takes the value $n$ $(n\geq 1)$ with probability $p_n$. When the LLM service provider sets up the maximum token limit $n_{max}$, users who require a greater number of output tokens will receive only up to $n_{max}$ tokens. Consequently, the expected mean of the output token numbers can be calculated as follows:
\begin{equation}
\label{eq:2.2}
E[n_{req}]=\sum\nolimits_{n=1}^{n_{max}{-}1} np_{n}+n_{max} \cdot (1{-}\sum\nolimits_{n=1}^{n_{max}{-}1} p_{n}), 
\end{equation}
where the service time $S$ is affected by the token limit. 

According to our experimental observation in Fig. \ref{fig2.3.1}, the inference latency of a single request is approximated by $S=an+c$ with $a$ and $c$ as constants. By basic probabilistic analyses, we obtain the following expressions:
\begin{eqnarray}
\label{eq:2.3}
E[n_{req}^{2}]&{=}&\sum\nolimits_{n=1}^{n_{max}-1} n^{2}p_{n}{+}n_{max}^{2}(1{-}\sum\nolimits_{n=1}^{n_{max}-1} p_{n}),\\
\label{eq:2.4}
E[S]&{=}&aE[n_{req}]+c,\\
\label{eq:2.5}
E[S^{2}]&{=}&E^{2}[S]+a^{2}(E[n_{req}^{2}]-E^{2}[n_{req}]).
\end{eqnarray}

Substituting Eq. \eqref{eq:2.5} into Eq. \eqref{eq:2.1} yields the explicit expression of the queueing delay with respect to the max token limit.

\subsection{Queuing Analysis with User Impatience}


Patience is often a scarce commodity among users when it comes to waiting for LLM inference services. After a certain period of waiting, users may decide to abandon the system. To delve deeper into the impact of the maximum token limit on the average inference delay, especially considering the impatience of users, we introduce a model for analysis. For the sake of mathematical simplicity, we assume that each incoming user is prepared to wait in the queue for a maximum of $\tau$ seconds before their request is processed. Users who choose to exit the inference system without having their requests processed are referred to as lost users.

Denote by $\pi(\tau)$ the fraction of the lost users in the long-run. Let $E[W_q]$ be the average queuing delay experienced by all the served and lost users, and let $E[W_{qs}]$ be the average queueing delay of the served users. Their approximate formulas have been studied in \cite{22} and are expressed as:
\begin{eqnarray}
\label{eq:2.6}
\pi(\tau)&=&(1-\zeta^2)\pi^{det}(\tau)+\zeta^2\pi^{exp}(\tau),\\
\label{eq:2.7}
E[W_q]&=&(1-\zeta^2)E[W_q^{det}]+\zeta^2E[W_q^{exp}],\\
\label{eq:2.8}
\zeta^2&=&\frac{E[S^2]-E^2[S]}{E^2[S]}.
\end{eqnarray}

This approximation requires $\zeta ^2$ to satisfy $0\leq \zeta^2 \leq 1$. $E[W_{qs}]$ can be determined using $\pi(\tau)$ and $E[W_q]$. It is not hard to understand that each lost user spends a time $\tau$ in the queue so that it follows: 
\begin{equation}
\label{eq:2.9}
E[W_q]=\tau \pi(\tau)+E[W_{qs}](1-\pi(\tau)).
\end{equation}


$\pi^{exp}(\tau)$ and $E[W_q^{exp}]$ represent the corresponding results with the exponentially distributed number of requested tokens. 
$\pi^{det}(\tau)$ and $E[W_q^{det}]$ represent the results in the special case of deterministic service requirements. Their calculation formulas are listed in \cite{22} and will not be repeated here.

Similarly, using Eq.\eqref{eq:2.5}, we obtain the expression of the mean queueing delay as a function of the max token limit. Intuitively, the larger token limit will lead to the higher service delay and more lost users. However, the number of requested tokens of user is intrinsic to his evaluation of LLM service. Configuring a median or even small token limit will decrease the user perceived content quality. Our model facilitates the choice of an optimal token limit that balances the generated content quality and the service delay.

\subsection{Optmization Model}

We formulate an optimization problem to find the optimal tradeoff between LLMs service quality and its queueing delay. 
The utility of the LLM service for a user is defined as a general function $u(n_{req})$, where $u(n_{req}):= 1$ if $ n_{req} \leq n_{max}$ and $u(n_{req}):= 1-\frac{n_{req}-n_{max}}{n_{req}}$ if $ n_{req} > n_{max}$. Denote by $V(n_{max})$ the objective function of the LLM service provider. When all the users are assumed to be sufficiently patient, $V_1(n_{max})$ is modeled as:
\begin{equation}
\begin{split}
\label{eq:2.14}
V_1(n_{max}) = \theta E[u| n_{max}] - (1-\theta) E[W(n_{max})],
\end{split} 
\end{equation}
where $E[u| n_{max}] {=} \sum\nolimits_{n=1}^{n_{max}} p_{n} {+}\sum\nolimits_{n=n_{max}{+}1}^{\infty} (1{-}\frac{n{-}n_{max}}{n})p_{n}$ and $\theta$ is the weighting factor. When the user impatience is considered, the LLM service provider incurs a fixed cost on each lost user. Here, we define $\ell$ as this cost. Then, the objective function $V_2(n_{max})$ is given by:
\begin{equation}
\begin{split}
\label{eq:2.15}
V_2(n_{max}) {=} \theta E[u| n_{max}] {+} (1{-}\theta) E[W(n_{max})] {-} \pi(n_{max}) \cdot \ell.
\end{split} 
\end{equation}
Our optimization problem is summarized as:
\begin{eqnarray}
\label{eq:2.117}
\max && V(n_{max}) \\
\label{eq:2.118}
s.t. && n_{max} \geq 1.
\end{eqnarray}

Here, Eq. \eqref{eq:2.117} is intended to optimize the LLM service level and users experience. To ensure rationality, Eq. \eqref{eq:2.118} imposes constraints on the max output token limit.

\section{Impact of Dynamic Batching}
In this section, we rigorously analyze the service delay for different batching methods including dynamic batching, fixed batching and elastic batching. 



\subsection{Basic Stochastic Model}

For GPU-based batching inference, Inoue \cite{18} models a single server dynamic batching service with infinite batch size. The requests arriving at the server follow the Poisson distribution with rate $\lambda$ and they can be processed simultaneously in a batch. The inference time of a batch depends on the batch size $b$, which is defined as $H^{[b]}$ $(b=1,2, \cdots)$. For non-AIGC DNN inference, a model inputs a vector whose size is the product of the fixed input dimension and the batch size. It  computes the output through a predefined sequence of operations so that the inference time increases linearly with the batch size:
\begin{equation}
\label{eq:2.16}
H^{[b]}=\alpha b+\beta.
\end{equation}

Let $\mu^{[b]}$ $(b=1,2, \cdots)$ denote the inference speed in a batch with size $b$, so that $\mu^{[b]}=\frac{b}{E[H^{[b]}]}$.

In this system, when the server becomes idle, all the pending requests in the queue are processed together in a batch. If the queue is empty, the server will wait until a request arrives. The size of the $i^{th}$ batch processed is defined as $B_i$ $(i=1,2, \cdots)$ and the number of requests arriving in the processing time of the $i^{th}$ batch is defined as $A_i$ $(i=0,1, \cdots)$. Then there exists:
\begin{equation}
\label{eq:2.18}
B_{i+1}=A_i+\mathbb{I}_{(A_i=0)},  \;\;\; i=0,1,\cdots,
\end{equation}
where $\mathbb{I}_{(\cdot)}$ denotes an indicator function. By mathematical derivations, the mean queuing delay $E[W]$ is bounded above by:
\begin{equation}
\begin{split}
\label{eq:2.19}
E[W]&\leq \frac{\lambda(\alpha+\beta)^2}{2(1-\lambda^2\alpha^2)}=:\phi_0(\lambda,\alpha,\beta),\\
E[W]& \leq \frac{\lambda\alpha\beta+\lambda\alpha^2+\beta}{2(1-\lambda^2\alpha^2)}=:\phi_1(\lambda,\alpha,\beta).\\
\phi(\lambda,\alpha,\beta)&:=\min(\phi_0(\lambda,\alpha,\beta), \phi_1(\lambda,\alpha,\beta)).
\end{split}
\end{equation}

\subsection{LLMs Dynamic Batching} 

Distinguished from traditional non-AIGC models, the inference time of a LLM request is not deterministic, but depending on the numbers of input and output tokens. 
Within a batch, different requests may have varying numbers of input (and output) tokens, necessitating padding to align with the longest input (and output). Consequently, the inference time of a batch is influenced not only by the batch size but also by the maximum input and output token sizes.
For requests involving long text inputs, such as article translation and summarization, users are typically less sensitive to queuing delays. Therefore, our focus here is primarily on short text inputs.  We suppose the number of input tokens follows the uniform distribution from $0$ to $m_1$ ($m_1 \geq 0$). 
For a batch $b$, the CDF of the max input token size $D_{max}$ is given by:
\begin{equation}
\label{eq:2.22}
\begin{split}
Pr(D_{max} \leq x|B=b)=(\frac{x}{m_1})^b, 0\leq x \leq m_1,
\end{split}
\end{equation}
then we can obtain the expectation of $D_{max}$ is $\frac{m_1b}{b+1}$.

Taking its upper bound, we can obtain a fixed value $m_1$, so that we can use the results in Fig. \ref{fig2.3.2}. The time to generate the first token is $k_1b+k_2$ and the time to generate others is $(k_3b+k_4)l$. $l$ denotes the max output token size in the batch, which contains the second output token to the last one. The inference time $H^{[b,l]}$ of the batch is given by:
\begin{equation}
\label{eq:2.24}
H^{[b,l]}=k_1b+k_2+k_3bl+k_4l,
\end{equation}
for some $k_1>0, k_2 \geq 0,k_3 \geq 0,k_4 \geq 0$. 

Then, we consider the different distributions of the number of output tokens, such as the uniform distribution, Gaussian distribution and so on.

\subsubsection{Uniform Distribution of Output Tokens}
we suppose that the number of output tokens $N$ follows the uniform distribution from $0$ to $m_2$ ($m_2 \geq 0$).

Let $L$ be a generic random variable of the max output token size in the batch, so that $E[L]=\frac{m_2b}{b+1}$ . For a batch with size $b$, the mean inference time is given by:
\begin{equation}
\label{eq:2.26}
H^{[b]}=  k_1b +k_2+k_3m_2 \frac{b^2}{b+1} + k_4m_2 \frac{b}{b+1}.
\end{equation}

Taking its upper bound, we can obtain,
\begin{equation}
\label{eq:2.27}
\begin{split}
H^{[b]}&\leq (k_1+k_3m_2)b+k_2+k_4m_2\\
&=\alpha_1 b+\beta_1.
\end{split}
\end{equation}
Eqs. \eqref{eq:2.27} and \eqref{eq:2.16} are the same in their forms, so that we can obtain the mean queuing delay according to Eq. \eqref{eq:2.19}.

\subsubsection{Truncated Gaussian Distribution of Output Tokens} we suppose the number of output tokens follows the truncated Gaussian distribution from 0 to $\infty$, where the mean is $u$ and the standard deviation is $\sigma$.

The PDF and CDF of the truncated output token size $N_{tr}$ ($N_{tr} \geq 0$) is given by:
\begin{eqnarray}
\label{eq:2.44}
f_{N_{tr}}(x)&=&\frac{\phi(\frac{x-u}{\sigma})}{\sigma (1-\Phi(\frac{-u}{\sigma}))},x\geq 0,\\
\label{eq:2.45}
F_{N_{tr}}(x)&=&\int_{0}^{x}  \frac{\phi(\frac{t-u}{\sigma})}{\sigma (1-\Phi(\frac{-u}{\sigma}))} dt,x\geq 0,
\end{eqnarray}
where $\phi(x)$ and $\Phi(x)$ are the PDF and CDF of the standard Gaussian distribution. Here, $L$ is the maximum order statistic regarding $N_{tr}$, and its expectation is given by:
\begin{equation}
\label{eq:2.46}
E[L]=\int_{0}^{\infty} bf_{N_{tr}}(x) (F_{N_{tr}}(x))^{b-1} dx.
\end{equation} 

\begin{figure}[!t]
\centering
\subfloat[]{\includegraphics[width=1.7in]{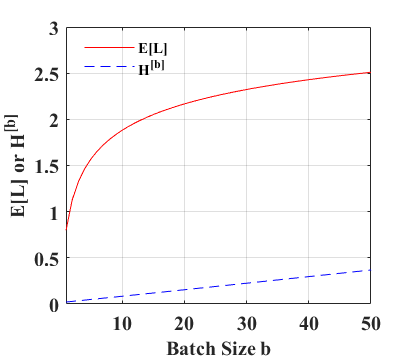}%
\label{fig3.1.1}}
\hfil
\subfloat[]{\includegraphics[width=1.7in]{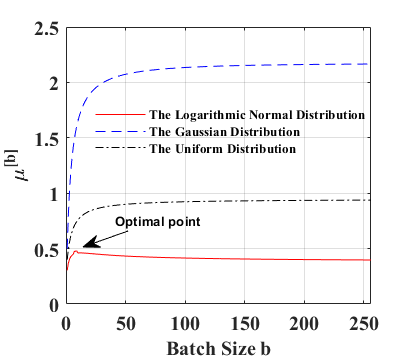}%
\label{fig3.1.2}}
\caption{The relationship between $E[L]$ (resp. $H^{[b]}$, $\mu^{[b]}$) and $b$.}
\label{fig3.1}
\end{figure}

We illustrate $E[L]$ and $H^{[b]}$ with regard to the batch size under the truncated standard Gaussian distribution in Fig. \ref{fig3.1.1}. Here $E[L]$ monotonically increases with the batch size and the rate of rise becomes smaller and smaller, which becomes a fixed value quickly. Therefore, $H^{[b]}$ increases approximately linearly with the batch size, so that we can also obtain the mean queuing delay according to Eq. \eqref{eq:2.19}. For other Gaussian distributions with small standard deviation, the results are similar.

\subsection{LLMs Fixed Batching}

In all distributions of the number of output tokens, it is not universally true that larger batch sizes lead to higher inference speeds, as depicted in Fig. \ref{fig3.1.2}. The figure contains the uniform distribution from 0 to 2000, the truncated Gaussian distribution with mean 800 and standard deviation 20 and the logarithmic normal distribution with log mean 7 and log standard deviation 0.7. The inference speed of a fixed batch size is given by:
\begin{equation}
\label{eq:2.48}
\mu^{[b]}=b/(k_1b+k_2+k_3bE[L]+k_4E[L]).
\end{equation}

We note that for a light-tailed distribution, $\mu^{[b]}$ increases with the
batch size $b$. However, with a heavy-tailed distribution, there exists an optimal batch size that maximizes the inference rate.


This understanding is straightforward: for a light-tailed distribution, $E[L]$ quickly reaches its maximum as the batch size increases, leading to a monotonically increasing function of $\mu^{[b]}$ with $b$. Conversely, with a heavy-tailed distribution, $E[L]$ progressively increases with the batch size. Initially, when the batch size is small, $E[L]$ and its impact on inference time are minimal, resulting in an increase in the inference rate. However, as the batch size grows, $E[L]$ becomes significantly larger and continues to increase, necessitating the padding of numerous tokens for many responses. This phenomenon slows down the inference rate.

Therefore, for the heavy-tailed distribution, we model the customer queuing delay under a fixed batch size. We model the system as an $M/D^b/1$ queuing system, where $D^b$ indicates deterministic bulk service with a fixed batch size $b$.
If the specified number of users, denoted as $b$, is present, a single server serves them simultaneously; otherwise, the server remains idle and waits until a total of $b$ users has accumulated before providing service. The batch inference time is a fixed number that contingent upon the batch size, where $H^{[b]}=k_1b+k_2+k_3bE[L]+k_4E[L]$. According to \cite{24}, the mean queueing delay is shown as follows.
\begin{equation}
\begin{split}
\label{eq:2.46}
E[W] &= \frac{1}{\lambda}(\frac{b-(b-\lambda H^{[b]})^2}{2(b-\lambda H^{[b]})}+\sum_{k=1}^{b-1} \frac{1}{1-Z_k}),\\
Z_k &= \sum_{m=1}^{20} c_mw_k^m, k=1,2,\cdots,b-1,\\
c_m &= \exp(\frac{-\lambda H^{[b]}m}{b})\frac{(\lambda H^{[b]}m)^{(m-1)}}{b^{(m-1)}m!},\\
w_k &= \exp(\frac{2\pi k}{b}i).
\end{split}
\end{equation}


Using MATLAB, we can determine the optimal batch size $b^*$ that minimizes $E[W]$. This approach allows servers to efficiently handle high arrival rates of requests. However, when the arrival rate $\lambda$ is low, users may experience delays until $b^*$ users are in the queue, resulting in wasted time. Therefore, dynamic batching with a maximum batch size $b_{max}$ becomes essential, where $b_{max}$ can be set to $b^*$.

\subsection{LLMs Elastic Batching}

Traditional batch processing of LLMs suffers from inefficiencies where shorter sequences are delayed by longer ones to ensure uniform token outputs, leading to computational waste. Yu and Joo et al. \cite{36} first proposed iteration-level scheduling, where each generated token is returned to the user immediately without waiting for all responses to complete. 

For modeling simplicity, we consider the case that when the batch of requests is completed can the next round of service be carried out. For a batch of requests, we let the replies that generate fewer tokens be returned to the customers in advance without padding. This also reduces the inference time for customers who need more output tokens, because as more and more replies are returned to the customers, the remaining batch size gets smaller and smaller.  

Assume a batch of requests with batch size $b$, the number of tokens generated from small to large is $n_1,n_2, \cdots,n_b$, where $n_i$ $(i=1,2,\cdots b)$ only contains the second output token to the last one. The latency to complete this batch of requests is the latency to complete the request which generates $n_b$ tokens. After $k_1b+k_2+(k_3b+k_4)n_1$ inference time, the reply which generates $n_1+1$ tokens is returned to the customer. At this time, the request which needs $n_2+1$ tokens only has $n_2-n_1$ tokens not generated. Then another $(k_3(b-1)+k_4)(n_2-n_1)$ inference time, the reply can be returned to the customer. And so on, the batch processing time can be expressed as follows.
\begin{equation}
\begin{split}   
\label{eq:2.47}
H^{[b,n_b]} &{=} k_1b{+}k_2{+}(k_3b{+}k_4)n_1{+}(k_3(b{-}1){+}k_4)(n_2{-}n_1)\\
&{+} \cdots{+}(k_31(b{-}(b{-}1)){+}k_4)(n_b{-}n_{b-1})\\
&{=}k_1b{+}k_2{+}k_3bE[N]{+}k_4n_b.
\end{split}
\end{equation}

The inference delay of Eq. \eqref{eq:2.47} is much smaller than that of Eq. \eqref{eq:2.24}. Similarly, the mean queuing delay can be calculated according to Eq. \eqref{eq:2.19}.

\section{Experimental Studies}

\subsection{Experiment Setup}


In this section, we analyze the inference latency curve based on output token size and batch size using a real LLM-based inference server on NVIDIA A100. Subsequently, we validate the practicality of our derived mathematical expressions through numerical experiments conducted using MATLAB's event-driven simulation. 
The base language models can only perform text continuation, i.e., generating the following text when given the preceding context, which cannot engage in conversational interaction. In contrast, an instruction-finetuned language models can understand user requests and provide appropriate responses. Therefore, three types of open-source instruction-finetuned language models are used in this experiment: LLaMA-2-7b-chat, Chinese-Alpaca-2-7b, and ChatGLM3-6b. For experimental data, we simulate virtual scenarios with output token sizes following various distributions.

\subsection{Impact of Token Limit}
\begin{figure*}[!t]
\centering
\subfloat[The mean queueing delay with patient users.]{\includegraphics[width=1.7in,height=1.2in]{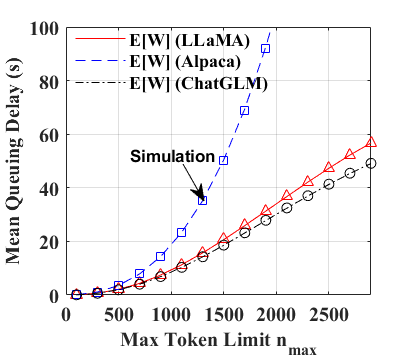}%
\label{fig4.4.1}}
\hfil
\subfloat[The mean queueing delay with impatient users.]{\includegraphics[width=1.7in,height=1.2in]{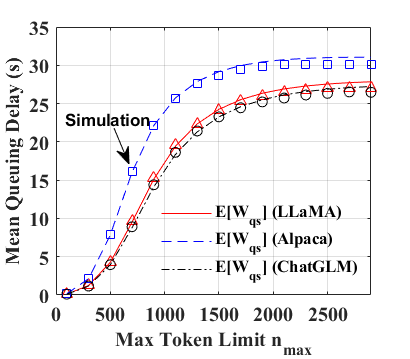}%
\label{fig4.4.2}}
\hfil
\subfloat[The fraction of the lost users because of impatience.]{\includegraphics[width=1.7in,height=1.2in]{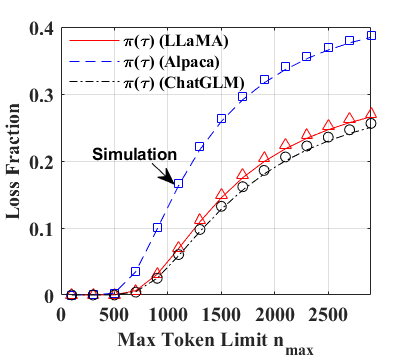}%
\label{fig4.4.3}}
\hfil
\subfloat[Tradeoff.]{\includegraphics[width=1.7in,height=1.2in]{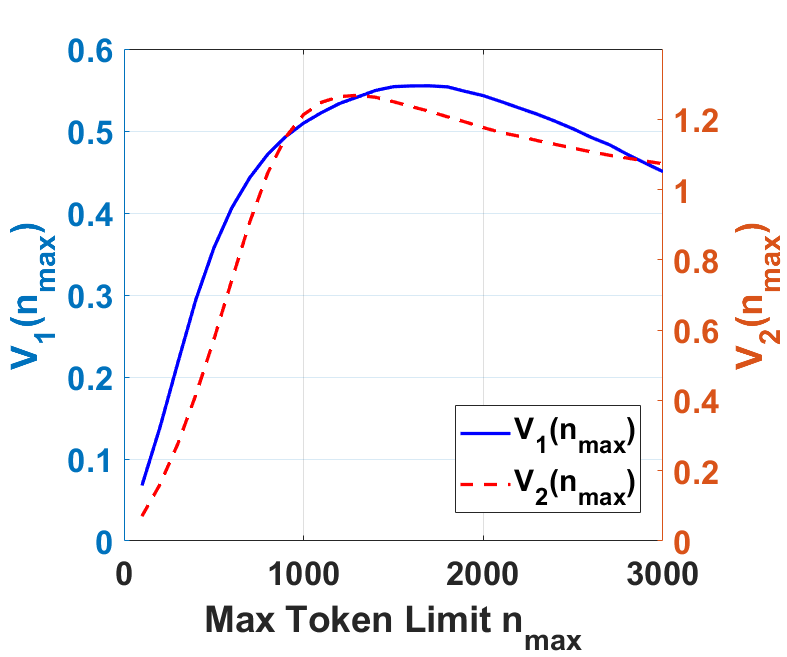}%
\label{fig4.4.4}}
\caption{The queueing results of max-limit clipping on A100 GPU.}
\label{fig4.4}
\end{figure*}

We assume that the output token size follows a logarithmic normal distribution with log mean 7 and log standard deviation 0.7. Fig.\ref{fig4.4.1}, Fig.\ref{fig4.4.2} and Fig.\ref{fig4.4.3} show the mean queuing delay $E[W]$, $E[W_{qs}]$ and loss fraction $\pi(\tau)$ with regard to the max token limit $n_{max}$. Among them, the red, blue, and black lines represent the numerical results obtained through mathematical equations, while the red triangles, blue squares, and black circles correspond to the simulation results. 
We randomly set the arrival rate such that $\rho<1$. 
The arrival rate $\lambda$ is 1/40 in Fig. \ref{fig4.4.1} and $\lambda$ is 1/25 and $\tau$ is 60 in Fig. \ref{fig4.4.2} and Fig.\ref{fig4.4.3}. We observed that each curve is accurately described by the provided expressions. When assuming all users are sufficiently patient, the mean queuing delay tends to increase with the maximum token limit. This is because a significant number of users requiring long responses enter the system, significantly slowing down LLMs inference. However, when considering user impatience, the mean queuing delay gradually stabilizes. This occurs because more users leave the system without being served, reducing the overall load over time. 

In order to provide better service, the optimal tradeoff on LLaMA-2-7b-chat is given in Fig. \ref{fig4.4.4}. Here, $\theta$ is 119/120 for the objective function $V_1(n_{max})$ and $\theta$ is 0.95 and $\ell$ is 4 for $V_2(n_{max})$. When all the users are assumed to be sufficiently patient, the optimal max token limit is 1600. The mean queuing delay is only 23s, which decreases by 58.93\% when $n_{max}$ is 3000. When the user impatience is considered, the optimal max token limit is 1300. Although the mean queuing delay does not decrease significantly, the fraction of the lost users is only 0.12, which decreases by 56.36\% when $n_{max}$ is 3000. This shows that more and more users can accept the service, which can bring greater profits to the LLMs platform. 
Among users, 70.53\% have an output token size less than 1600, while 59\% have an output token size less than 1300. This indicates that enforcing a maximum output token limit on a small fraction of inference requests can better serve users.

\begin{figure*}[!t]
\centering
\subfloat[LLaMA-2-7b-chat]{\includegraphics[width=2.3in,height=1.7in]{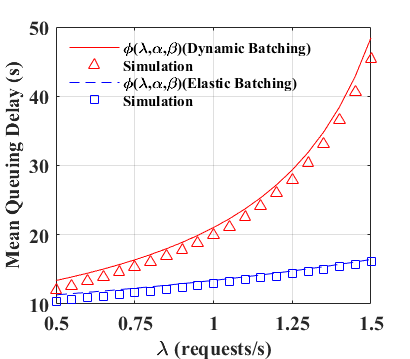}%
\label{fig4.12.1}}
\hfil
\subfloat[Chinese-Alpaca-2-7b]{\includegraphics[width=2.3in,height=1.7in]{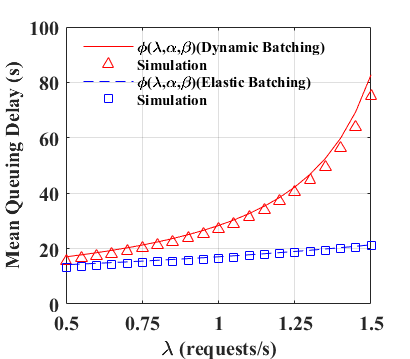}%
\label{fig4.12.2}}
\hfil
\subfloat[ChatGLM3-6b]{\includegraphics[width=2.3in,height=1.7in]{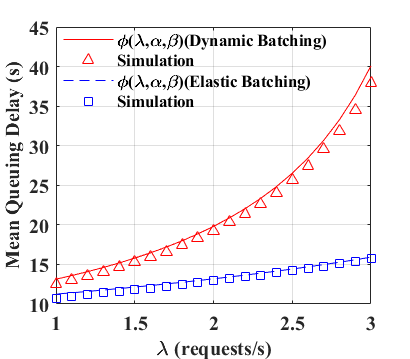}%
\label{fig4.12.3}}
\caption{The mean queueing delay using dynamic batching and elastic batching on A100 GPU.}
\label{fig4.12}
\end{figure*}

\subsection{Impact of Batching Strategies}

Fig. \ref{fig4.12} illustrates simulation results for the mean queueing delay and its upper bound under traditional dynamic batching and elastic batching. We assume the output token size follows a uniform distribution ranging from 0 to 1000.
The upper bound closely approximates the exact curve of the mean queueing delay. Specifically, elastic batching exhibits lower mean queueing delays compared to traditional dynamic batching, and this advantage increases with the arrival rate. 
Intuitively, as the arrival rate increases, more customers enter the LLMs system, leading to larger batch sizes. This results in larger maximum output token sizes within the batch, causing numerous short responses to be padded with many tokens and thereby significantly increasing the mean queueing delay. However, elastic batching effectively mitigates this issue.

\begin{figure}[!t]
\centering
\subfloat[$\lambda=0.43$]{\includegraphics[width=1.7in]{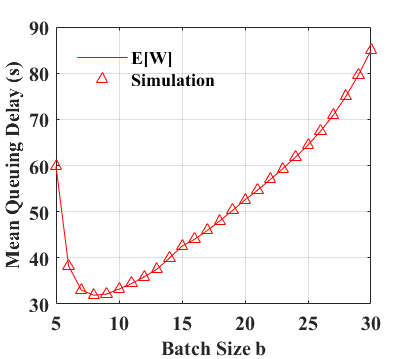}%
\label{fig4.13.1}}
\hfil
\subfloat[Dynamic batching with max batch size and elastic batching]{\includegraphics[width=1.7in]{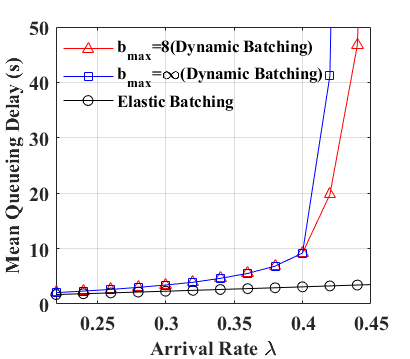}%
\label{fig4.13.2}}
\caption{The mean queueing delay of different batching inference for LLaMA-2-7b-chat on A100 GPU.}
\label{fig4.13}
\end{figure}

If the number of output tokens follows a heavy-tailed distribution, the relationship between batch size and queueing delay is not straightforwardly that larger batch sizes lead to reduced queueing delays, as discussed earlier. Therefore, traditional dynamic batching, which processes all requests in the queue together, may not be suitable. Instead, it becomes crucial to determine the optimal batch size.
Fig. \ref{fig4.13.1} presents simulation results of the mean queueing delay, alongside its mathematical formulas given in Eq. \eqref{eq:2.46}. The number of output tokens follows a logarithmic normal distribution with a log mean of 7 and a log standard deviation of 0.7. Furthermore, based on calculations, the queueing delay under Eq. \eqref{eq:2.19} is estimated to be 125 seconds when the request arrival rate is 0.43. This delay is significantly higher than that observed with an optimal batch size of 8. 

When the arrival rate of requests is large, the mean queueing delay of optimal batch size is less than traditional dynamic batching. However, when $\lambda$ is small, the arriving users can not be immediately served until $b^*$ users in the queue, so that the mean queueing delay is large than traditional dynamic batching. Therefore, dynamic batching with max batch size $b_{max}$ is necessary, which is given in Fig. \ref{fig4.13.2}. The output token distribution is the same as in Fig. \ref{fig4.13.1}. When $\lambda$ is small, as long as $b_{max}$ is larger than the average number of requests in the queue, we can obtain the same low queuing delay. When $\lambda$ is large, we find that the mean queuing delay of dynamic batching with $b_{max}=8$ is much less than infinite batch size. Therefore, the system of dynamic batching with $b_{max}=b^*$ can provide better service. 
At the same time, we observe that elastic batching is better than traditional dynamic batching even for heavy-tailed distribution, as shown in Fig. \ref{fig4.13.2}.

\section{Related Work}
\label{sec:related}
Due to high latency for LLMs, numerous techniques have been proposed to reduce inference time. Liu and Wang et al. \cite{26} show that contextual sparsity exists, which uses the current input to dynamically select part of the network parameters for inference instead of using all of them. In \cite{28}, Zheng and Ren et al. accurately perceive and predict the response length, so that queries with similar response length can be gathers into micro-batches. Xiao and Lin et al. \cite{29} propose an INT8 quantization of both weights and activations for all the matrix multiplications in LLMs. However, they do not improve the service of LLMs from the queueing theoretic perspective.
In \cite{31}, Li and Zheng et al. only analyze the queuing theory of LLMs from the perspective of parallelization.
The impact of parameter configuration on the inference performance has been studied, such as the prompt in \cite{32,33} and temperature in \cite{34}. However, there is no detailed mathematical derivation of how to choose the optimal max token limit.

There has been many works analyzing the batch-service queues. Works in \cite{8,9,10,11} assume that the processing time has nothing to do with the batch size. 
Inoue \cite{18} assumes the batch processing time linearly increases with the batch size so that derives a closed-form upper bound of the mean queuing delay. However, the batch inference time of LLMs is related to not only the batch size, but also the maximum input and output token sizes in this batch.


\section{Conclusion}
\label{sec:conclusion}

This paper models the LLM inference delay from the perspective of queueing theory, and concentrates on the distribution of the size of output tokens. For the separated task inference, we formulate M/G/1 models to derive the queueing delay, showing that under the heavy-tailed output token distribution, a very small fraction of tasks with large output token size significantly increase the queueing delay, and even drive a number of impatient users to leave the system. Our model shows that enforcing the max-token limit can reduce the queueing delay and the user loss rate simultaneously. For different batch processing methods, we develop a suit of bulk queue models, and derive the upper bounds of the queueing delay. We observe that when the output toke size conforms to light-tailed distributions, dynamic batching without the batch size limit outperforms the fixed batching. Conversely, when dealing with heavy-tailed distributions, dynamic batching with a finite batch size is more effective. Regardless of which distribution the output token size obeys, the elastic batching without intra-batch waiting has a minimum queueing delay.

\bibliographystyle{IEEEtran}
\bibliography{reference}

\end{document}